\documentclass[aps,prl,reprint,groupedaddress]{revtex4-1}
\usepackage{epsfig}
\usepackage{graphicx}

\begin{document}

\title{BZT: 
%
a 
soft pseudo-spin glass}

\author{David Sherrington}
\email[]{D.Sherrington1@physics.ox.ac.uk}

\affiliation{ Rudolf Peierls Centre for Theoretical Physics, University of Oxford, 
1 Keble Road, Oxford OX1 3NP, UK\\
Santa Fe Institute, 1399 Hyde Park Rd., Santa Fe, NM 87501} 

\date{24 Oct 2013}

\begin{abstract}

In an attempt to understand the origin of relaxor ferroelectricity, it is shown that interesting  behaviour of the onset of non-ergodicity and of precursor nanodomains, found in first principles simulations of the 
relaxor alloy $\mathrm {Ba(Zr}_{1-x}\mathrm{Ti}_{x}\mathrm{)O}_3$, can 
easily
be understood 
%
within
a simple mapping to 
%
a
soft pseudo-spin glass. 
\end{abstract}
\maketitle

For several years there has been much interest in 
relaxor ferroelectric
alloys based on the the generic 
pure ionic
perovskite form $\mathrm{ABO}_{3}$, where  A, B, O have charges +2, +4 and -2, 
but with the single B-type ion replaced by random mixtures of ${\rm{B', B''}}$
\cite{Cross,Cowley, foot-A}.
%
 %
%
They
exhibit (i) frequency-dependent peaks in their dielectric susceptibilities as a function of temperature but without any macroscopic polarization in the absence of applied fields and (ii)
higher temperature manifestations of nano-scale polar domains \cite{Egami}. 
They have proven to be of significant application value but there is no universally accepted understanding of the origin of their behavior. 
The present objective  is to provide such understanding within the context of 
a recently recognised system, 
employing only minimal modeling and simple mappings
and without the need to posit 
{\it{a priori}}
random bonds or random fields.
%


The originally discovered  \cite{Smolenskii} and most studied relaxor is  $\mathrm {Pb(Mg}_{1/3}\mathrm{Nb}_{2/3}\mathrm{)O}_3$ (PMN). It exhibits the features mentioned above, as well as non-ergodicity \cite{Kleemann_FC_ZFC, Levstik}  beneath a temperature comparable with that of the finite-frequency susceptibility peaks. However, it
is complicated by the fact that Mg and Nb are not isovalent, 
giving rise to perturbing extra charges  and hence random fields.  By contrast, in 
$\mathrm {Ba(Zr}_{1-x}\mathrm{Ti}_{x}\mathrm{)O}_3$ (BZT)
 Zr and Ti  are isovalent, of charge +4. Yet it still exhibits characteristic relaxor features \cite{Maiti, Shvartsman, Kleemann-Miga} for a range of relative (Zr:Ti) concentations.

A recent first principles and Monte Carlo computer simulation study of 
BZT \cite{Akbarzadeh2012}
has demonstrated an ergodicity-breaking phase transition at which a separation onsets between dielectric susceptibilities measured under 
different protocols and also has exhibited nano-domains above this transition temperature. 
This communication provides  a physical explanation of this transition as the onset of a soft spin glass-like state, extends the analogy to explain the more general phase structure of BZT and demonstrates  an 
expected 
origin of the observed nano-domains. 

$\rm{ BaZrO_3}$ and $\rm{ BaTiO_3}$ are ${\rm ABO_3}$ ionic crystals with positive charges on the Ba, Zr and Ti ions and negative
charges on the O ions. Their equilibrium structures correspond to minimizing their free energies under the resultant competing 
(spatially frustrated) 
interactions. At high temperatures both have simple perovskite stucture but at low temperature $\rm{ BaTiO_3}$ 
transforms to a ferroelectric through spontaneous coherent displacement of the Ti ions; $\rm{ BaZrO_3}$ remains paraelectric as the temperature is lowered. The particular current interest is in alloys in which the B sites
are occupied randomly by Zr or Ti.

Akbarzadeh et al.\cite{Akbarzadeh2012} studied the alloy system   with equal concentrations of Zr and Ti,  allowing for 
displacements of all the ions (in a finite-size simulation) and using parameters obtained from 
first-principles computer modeling of small cells.  They examined the susceptibilities 
measured (i) by directly observing the polarization when cooled in a small applied field and (ii) from 
the correlation function in the absence of an applied field, using  the conventional equilibrium statistical physics relationship 
between response and correlation. These measurements  increased with reducing temperature and roughly coincided  above a 
characteristic temperature, $T_f$, but  started to diverge significantly
from one another at  this temperature, with the directly evaluated susceptibility exhibiting a plateau beneath  it while that
 determined from the correlations decreased, giving a cusp at $T_f$. This is precisely what is expected from  mapping to a pseudo-spin glass. 

As noted and utilised in \cite{Akbarzadeh2012}, a crucial difference between systems with Zr or Ti at a B site lies in the strength of the effective local restoring forces associated with displacements of the 
 ions from their positions in the pure matrix; these are weak for Ti, permitting the low-temperature ferroelectric distortion 
observed in $\rm{ BaTiO_3}$, whereas in $\rm{ BaZrO_3}$ the Zr restoring force is much stronger and prevents macroscopic global distorsion even to zero temperature.  

Akbarzadeh et al. \cite{foot-Bellaiche}  modeled the alloy 
in terms of local mode variables centred on B-sites,
including averaged  inter-site interaction terms, simple local restoring-force terms of strengths corresponding to the appropriate local B-site occupation and associated random fields and random strains. However, they found that their results are essentially unaffected by the random field and random strain terms and hence these will be ignored from the outset here. For conceptual purposes the modeling can be simplified further by absorbing the effects of the Ba and O ions into an effective system involving only the B-site ions.
Ignoring  any local anisotropy for illustrative simplicity, 
one is then left with a model characterized by an effective Hamiltonian
\begin{equation}
H=\sum_{i} \{{\kappa_{i}{|\bf{u_{i}}|^2}} + {\lambda_{i}}{|\bf{u_{i}}}]^4\}     +
{\sum_{ij} H^{avg}_{int}({\bf{u}}_{i}, {\bf{u}}_{j},  {\bf{R}}_{ij})}
\label{eq:H2}
\end{equation}
where the sites $\{i\}$  are occupied randomly by Ti or Zr according to the admixture concentration, with corresponding  
$\kappa,\lambda$. 
$H_{int}$ represents interactions between $\{{\bf{u}}\}$ at different sites; the superscript $\{avg\}$ indicates that, as in \cite{Akbarzadeh2012}, 
effects of the randomness are averaged
and
%
%
details
of quenched randomness in $H_{int}$
are
%
ignored. 
The zero-temperature phase structure is given by minimising $H$ with respect to the $\{\bf{u}_{i}\}$. 

Considering first a pure system, the sign of $\kappa$ determines whether this Hamiltonian can, in principle, exhibit displacive or order-disorder transitions, with positive $\kappa$ being displacive and the true order-disorder limit corresponding to strongly negative  $\kappa$, in each case with $\lambda$ positive. Within mean field theory, in the order-disorder case there will always be a transtion to an ordered phase as the temperature is lowered from the high temperature paramagnetic phase, whereas in the displacive case a minimal strength of bootstrapping binding energy gain from $H_{int}$, through self-consistent displacements, is needed to overcome the local penalty from the $\kappa$ term.
For 
${\rm{ BaTiO}}_3$  $\kappa^{\rm{Ti}}$ is small enough 
%
to permit
ferroelastic, and hence ferroelectric, order  
being favored 
in this case\cite{foot-Zhong1}.  By contrast, $\kappa^{\rm{Zr}}$ is too large for self-consistent displacive order and only para-electricity $\{u_{i}=0\}$ is possible at all temperatures for ${\rm{ BaZrO}}_3$.

Turning now to the alloy and noting that the large $\kappa ^{\rm{Zr}}$  implies that all the sites ${\{i\}}$ occupied by Zr atoms have $u_{i}=0$ and hence may be ignored, one is left with the effective Hamiltonian 
\begin{eqnarray}
H_{eff} 
 = 
\sum_{i(Ti)}   \{\kappa^{\rm{Ti}}   {|{\bf{u}}_{i}|}^2 
+   {\lambda^{\rm{Ti}}} {|\bf{u_{i}}|^4}\}   \nonumber
 \\
 + 
{\sum_{ij(Ti)} } H_{int}({\bf{u}}_{i}, {\bf{u}}_{j},  {\bf{R}}_{ij}).
\label{eq:Heff}
\end{eqnarray}
with sums now restricted to  B-sites occupied by Ti ions.

The  fact that experimentally the low temperature state of $\rm{ BaTiO_3}$ is ferroelectric shows that the dominant interaction in $H_{int}$ is ferroelectric. However, there are  both ferroelectric and anti-ferroelectric contributions at different separations \cite{Zhong}
\cite{foot-Zhong}.

This model is now recognizable as 
%
a
 soft pseudo-spin  
analog 
of canonical 
experimental spin glass systems \cite{Mydosh}, such as ${\rm{Au}}_{1-x}{\rm{Fe}}_{x}$ or ${\rm{Eu}}_{x}{\rm{Sr}}_{(1-x)}{\rm{S}}$,  whose Hamiltonians may be written as
\begin{equation}
H=-\sum_{ij(Mag)} J({\bf{R}}_{ij}){\bf{S}}_{i}.{\bf{S}}_{j}
\label{eq:Hsg}
\end{equation}
where the ${\bf{S}}_{i}$ are hard spins 
\cite{foot-hard-spin}, 
$J({\bf{R}})$ is a translationally-invariant but
spatially-frustrated exchange interaction and the sum is restricted to sites occupied by magnetic atoms
\cite{foot-AuFe}. For large $x$, 
high concentrations of magnetic atoms, 
these systems are periodically magnetically-ordered but for lower concentrations of magnetic atoms a non-periodic non-ergodic but still cooperative spin-glass phase results
\cite{foot-frustration}.

With this identification it becomes clear that  within some intermediate concentration range  $x_c  > x > x_p$ of Ti on the B sites in the alloy ${\mathrm{BaZr}}_{(1-x)}{\mathrm{Ti}}_{x}\mathrm{O}_3$,  there will be a pseudo-spin glass transition at a critical temperature $T_g(x)$, marking the onset of non-ergodicity and preparation-dependence, the Zero-Field-Cooled (ZFC) susceptibility peaking and the Field-Cooled (FC) susceptibility `freezing' \cite{Nagata79}.
Given that the FC susceptibility essentially measures a full Gibbs average over all pure states while the ZFC essentially measures the susceptibility restricted to a single pure macrostate \cite{Mezard} \cite{Parisi}, this explains the corresponding observations of Akbarzadeh et al. \cite{Akbarzadeh2012}, with their $T_f$ identified as $T_g$, their  $x=0.5$ being within this relaxor/pseudo-spin-glass concentration range \cite{Maiti} and FC and ZFC corresponding to the  two different susceptibility measurements they made
\cite{foot-relaxor}. 

For $x>x_c$ the transition is to ferroelectricity at a $T_{c}(x)$ that increases with $x$, reaching the pure $\rm{ BaTiO_3}$ value at $x=1$ . As $x$ is decreased below $x_c$,  $T_g$ is expected also to decrease with $x$, but initially less quickly, 
 until a further critical concentration $x_p$ beyond which only paraelectricity exists as a thermodynamic phase; thus we have the sequence  with increasing Ti concentration ($x$)
paraelectric$\rightarrow$relaxor$\rightarrow$ferroelectric
for $0 \leq  x_{p} \leq x_{c} \leq 1$, 
\cite{names}, 
in accord with experiments \cite{Maiti,Shvartsman}.

As noted earlier,  the best known signature of relaxors is the feature of frequency-dependent  peaks in the susceptibility as a function of temperature, with the peak temperature increasing with increasing frequency \cite{Smolenskii} 
\cite{foot-peak}.
 It is 
observed experimentally for BZT 
\cite{Maiti, Shvartsman, Kleemann-Miga}. A similar frequency-dependent peaking  is also a 
well-known feature of spin glasses; 
see {\it{e.g.}} \cite{Tholence, Wassermann}. In spin glasses it is also well-known that the peak temperature tends in the zero-frequency limit to that of the onset of non-ergodicity as measured by deviation of the FC and ZFC susceptibilities. Hence the mapping above would also lead one to expect this famous relaxor signature.

There has been much interest in the precursor observation (or interpretation) of `nano-domains' in relaxors  and these were also seen in the simulations of Akbarzadeh et al.\cite{Akbarzadeh2012}, as well as in experiments \cite{Maiti}. They too can be understood  from the above `induced-moment' soft pseudo-spin modeling, as corresponding to longish-lived `local moments' on statistically occurring clusters of Ti ions. To see this $H$ and $H_{eff}$  may be re-interpreted as Ginzburg-Landau free energies with their parameters renormalized as a function of temperature. The effective `local nano-domains' are given by minimization with respect to the $\bf{u_{i}}$, yielding values given in simple mean field theory by the self-consistent solution of
\begin{equation}
{\tilde\kappa_{i}} {\bf{u_{i}}} 
+2{\tilde\lambda_{i}} {\bf{u_{i}}} {|\bf{u_{i}}|}^2
+ {\sum_{j} \partial { \tilde H_{int}({\bf{u}}_{i}, {\bf{u}}_{j},  {\bf{R}}_{ij})}/\partial{\bf{u}}_{i} }=0,
\label{self-cons}
\end{equation}
with all the terms effectively temperature-renormalized but with the most important conceptual feature that the $\{\tilde\kappa\}$ increase with increasing temperature relative to the interaction term.
This equation is closely analogous to that for a mean field theory of cluster moment formation in transition metal alloys introduced in \cite{Mihill} and, similarly to that case, the formation of local nanodomains is relatable to an Anderson localization model \cite{Anderson1958, Anderson1978}.

Simplifying for illustrative purposes to a simple scalar analog of eqn.($\ref{self-cons}$) we consider 
\begin{equation}
\tilde\kappa_{i} u_{i} 
+2 \tilde\lambda_{i} u_{i}^3
- \sum_{j} \tilde J_{ij} u_{j} =0.
\label{simplified_sc}
\end{equation}
and compare it with an Anderson-type eigen-equation
\begin{equation}
\tilde\kappa_{i} \phi_{i} 
-\sum_{j} \tilde J_{ij} \phi_{j} =E\phi_{i}.
\label{Anderson_equiv}
\end{equation}

Non-zero $u$ solutions to eqn.($\ref{simplified_sc}$) correspond to solutions of eqn.($\ref{Anderson_equiv}$) with $E<0$. However, solutions to eqn. ($\ref{Anderson_equiv}$) with quenched $\kappa$-disorder can be either localized or extended; localized states at the extremities of the band of eigenstates separated from a region of extended states by lower $E_{m_L}(x,T)$ and upper  $E_{m_U}(x,T)$ `mobility edges'. Note that  the density of states and the mobility edges are temperature-dependent through the renormalization of the $\tilde\lambda$ and $\tilde J$, decreasing with decreasing $T$.   Thus, the onset of mean-field `cluster moments', observable on finite timescales as nano-domains, is given by the onset of  solutions $E \leq 0$ to eqn.($\ref{Anderson_equiv}$), while the true thermodynamic transition, which requires an extended state,   occurs only  when the mobility edge $E_{m_L}(x,T)$ becomes zero. 

While in the usual electronic Anderson situation the $\{\tilde J_{ij}\}\ge0$  so that extended states are ferroelectric, in the present frustrated case with $\{\tilde J_{ij}\}$ of both signs the extended states can also be spin-glass-like \cite{foot-without} for finite $x$. This leads to the expectation of a true thermodynamic ferroelectric phase structure  as temperature is lowered at high $x$,  passing over to transition to a spin glass phase as $x$ is reduced to a critical $x_c$ and eventually beneath $x_p$ exhibiting paraelectric behavior only, but also with higher temperature non-equilibrium nanodomain precursors for all  $0<x<1$, the size of the precursor region reducing to zero as the pure limits are approached
\cite{foot-nanodomains}.

We might also note that in the Anderson analogy above quasi-frozen nano-regions need not necessarily be internally ferroelectric and indeed deviations from collinearity were observed in the simulations of \cite{Akbarzadeh2012}.

The concept of polar nano-regions (PNRs) interacting among themselves and eventually freezing cooperatively macroscopically can, in principle, be given substance by defining nano-moments in terms of negative eigenvalue eigenfunctions of eqn.(\ref{Anderson_equiv}), introducing them into an expanded partition function by adding them as variables with delta functions ensuring their identification and then integrating out the original variables \cite{foot-PNR}.

Note that neither random fields nor random interactions were posited above \cite{foot-Akb, foot-however}. However, $H_{eff}$ can be mapped into a random-bond model 
 %
\begin{equation}
H_{eff}^{EA} = \sum_{l}{\kappa{u_{i}^2} + \lambda {u_{i}}^4}   
+ \sum_{lm} J_{lm}u_{l}u_{m} 
\label{HEASS}
\end{equation}
where now the $l$ are relabelled Ti sites, $\kappa$ and $\lambda$ are site-independent and all the randomness is in the $\{J_{lm}\}$. In a precise mapping the $\{J_{lm}\}$ code the  spatial distribution of Ti ions. Following the conceptualization introduced by Edwards and Anderson \cite{EA} that the important physics of spin glases is maintained as long as one retains frustration and quenched disorder, one would expect that a further assumption of independent randomness of $\{J_{lm}\}$  would maintain the crucial physics.  However, to allow for the transition between ferroelectric and relaxor phases with $x$, the $\{J_{lm}\}$ distribution should have a tunable ($x$-dependent) mean \cite{SS,foot_SS}. 




These analogies also suggest that, within the relevant intermediate range of $x$,  BZT  should exhibit other behaviors corresponding to those known for spin glasses,
%
not only at the onset (where the susceptibility peaks), but also within the relaxor phase.
%
Similar behavior  and explanation might also be anticipated in other isovalent alloys of a frustrated displacive (or mixed displacive-weak order-disorder) ferroelectric (or antiferoelectric) and an appropriate  paraelectric partner 
\cite{foot-PMN}. The corresponding analogy between hard dipolar (strong order-disorder) and other orientational glasses and hard spin glasses has long been recognised \cite{Brout}; for reviews see \cite{Hochli} \cite{Binder}.

The modeling of eqn.($\ref{self-cons}$) is of course only mean-field and so misses both thermal fluctuation effects and dynamics. However a similar extended simple modelling based on disorder only in local restoring terms and a spatially frustrated periodic interaction could in principle be extended to treat these. 

Finally, it should be emphasised that the discussion above is minimal, a skeleton modeling  to expose the physical core. More `flesh' is needed for the whole body,


\section*{Acknowledgements}

The author thanks Prof. Rasa Pirc for drawing his attention to \cite{Akbarzadeh2012} and for helpful comments on a first draft, Profs. Wolfgang Kleemann and Laurent Bellaiche for comments, information and references, and the Leverhulme Trust for the award of an Emeritus Fellowship.


\begin{thebibliography}{99}
\bibitem{foot-A}{In some cases A is also substitutionally alloyed. }
\bibitem{foot-Zhong1}{In fact, detailed theoretical/numerical studies of ${\rm{ BaTiO}}_3$ \cite{Zhong}, of greater complexity than employed here,  have exhibited a ferroelectric state that has features of both displacive and order-disorder type, as expected for an effective $\kappa$ that is small and negative.}
\bibitem{foot-Zhong}{This is expected from the competing interactions in the ionic crystal, evidenced by the existence of  $\rm{ABO}_3$ antiferroelectrics, demonstrated in \cite{Zhong}, and clearly needed for the spin-glass like behavior discussed here.}
\bibitem{foot-AuFe}{In these alloys the Fe and Eu are the magnetic atoms.}
\bibitem{foot-frustration}{Note that the key necessary, but not automatically sufficient, ingredients for  spin glass behavior are frustration and quenched disorder. In the case of metallic spin glasses, such as AuFe, the exchange interaction is long-ranged and oscillatory (RKKY) whereas for a semiconductor, such as EuSrS, the range is shorter
%
and of different origin,
%
 but still greater than nearest neighbour and frustrated. For a brief theoretical perspective see \cite{DS-perspective}.}
\bibitem{foot-relaxor}{Note that the expression `relaxor' is used both for a phase and also to label materials that exhibit this phase. Here the phase is being understood to be that in which ergodicity-breaking is observed, measured quantities depending on the protocol used to observe them.}
\bibitem{foot-nanodomains}{A similar picture of nanodomains as corresponding to localized states above the relaxor phase has been 
suggested before \cite{Nambu} but for quenched randomness in the  $\{J_{ij}\}$ with $P(J_{ij}) = P(-J_{ij})$ and without the present simple mapping as the justification. As an example of the observation of  quasi-localized nanodomains, corresponding to Anderson eigenstates below the mobility edge but without the explanation above,  see \cite{Levin}.}
\bibitem{foot-PNR}{One might note that these nano-moments need not necessarily all be internally collinear nor even have a finite overall polarization.}
\bibitem{foot-however}{This does not mean that random fields cannot be relevant, only that they are not needed for the present explanation.}
\bibitem{foot_SS}{To emulate crudely the resultant interaction distribution and yield a transition sequence with increasing $x \in [0,1]$
paraelectric$\rightarrow$relaxor$\rightarrow$ferroelectric
one might take the mean
$ \overline{J_{ij}} = J_{0}\sim {x}$ and the standard deviation 
$[\overline {{(J_{ij}}^2}- {\overline {J_{ij}} ^2}]^{1/2}=J \sim {(x(1-x)})^{1/2}$.}
\bibitem{foot-PMN}{The classic relaxor alloy PMN ($\mathrm {Pb(Mg}_{1/3}\mathrm{Nb}_{2/3}\mathrm{)O}_3$), however, is not isovalent and consequently has additional interactions, together with random fields, due to the excess charges of the Mg (-2) and Nb(+1), as compared with Ti in the ferroelectric template $\mathrm {Pb}\mathrm{Ti}{\mathrm{O}}_3$. Hence it requires further consideration.}
\end{thebibliography}
\end{document}